\documentstyle[11pt,psfig,aaspp4]{article}
\begin{document}

\title{ MUELLER MATRIX PARAMETERS FOR RADIO TELESCOPES AND THEIR
OBSERVATIONAL DETERMINATION}

\author{Carl Heiles}

\affil {Astronomy Department, University of California,
    Berkeley, CA 94720-3411; cheiles@astron.berkeley.edu}

\author{Phil Perillat, Michael Nolan, Duncan Lorimer, Ramesh Bhat,
Tapasi Ghosh, Murray Lewis, Karen O'Neil, Chris Salter,
Snezana Stanimirovic}
\affil {Arecibo Observatory, Arecibo, PR 00613; email addresses
available at {\it www.naic.edu}}

\begin{abstract}

	Modern digital crosscorrelators permit the simultaneous
measurement of all four Stokes parameters.  However, the results must be
calibrated to correct for the polarization transfer function of the
receiving system.  The transfer function for any device can be expressed
by its Mueller matrix.  We express the matrix elements in terms of
fundamental system parameters that describe the voltage transfer
functions (known as the Jones matrix) of the various system devices in
physical terms and thus provide a means for comparing with engineering
calculations and investigating the effects of design changes.  We
describe how to determine these parameters with astronomical
observations.  We illustrate the method by applying it to some of the
receivers at the Arecibo Observatory. 

\end{abstract}


\section{INTRODUCTION}  \label{introduction}

	The polarization response of a single-dish radio telescope is
usually described in terms of the native polarization of the feed.  For
example, a dual-linear polarized feed provides two outputs
(``channels'') which are orthogonal linear polarizations; the sum is
Stokes $I$ and the difference Stokes $Q$.  Correlating the two outputs
with zero phase difference provides the other linearly polarized Stokes
parameter $U$, and correlating with $90^\circ$ phase difference provides
the circularly polarized Stokes parameter $V$.  Similarly, with a
dual-circular feed the sum and difference provide $I$ and $V$, and the
correlated outputs provide $Q$ and $U$.  However, these statements are
only approximate.  Feeds are almost never perfect, so their
polarizations are only approximately linear or circular; moreover, some
feeds, such as turnstile junctions, can have two orthogonal
polarizations with arbitrary ellipticity that is frequency-dependent. 
And generally speaking, no feed has two outputs that are perfectly
orthogonal. 

	The feed's response is modified by the electronics system, which
introduces its own relative gain and phase differences between the two
channels.  These must be calibrated relatively frequently because they
can change with time.  For example, if the feed is perfect and the
relative gain differs, then the difference between the two channels is
nonzero for an unpolarized source, making the source appear to be
polarized.  For a perfect linearly polarized feed, if the relative phase
of the two channels differs, then a linearly polarized source appears to
be circularly polarized.  Thus two quantities, the relative gain and
phase, must be calibrated.  This is most effectively done by injection
of a correlated noise source into the two feed outputs or,
alternatively, by radiating a noise source into the feed with a
polarization that provides equal amplitudes in the feed outputs. 

	The modification of the four Stokes parameters by these system
components is most generally described by a $4 \times 4$ matrix.  This
matrix is known as the Mueller matrix (Tinbergen 1996).  Along with
every Mueller matrix goes a Jones matrix, which describes the transfer
function for the voltages.  We express the Jones matrix elements in
terms of complex voltage coupling coefficients.  We consider three such
matrices: one describes the polarization state of the feed (circular,
elliptical, or linear); one the nonorthogonality of the polarized
outputs; and one the relative gain and phase of the electronics system. 
We calculate the associated Mueller matrices and describe how to solve
for the coupling coefficients from astronomical measurements.  We
illustrate the method by applying it to some of the systems at Arecibo
Observatory\footnote{The Arecibo Observatory is part of the
National Astronomy and Ionosphere Center, which is operated by Cornell
University under a cooperative agreement with the National Science
Foundation}. 

 	We begin by reviewing the basic theoretical concepts and
developing the structure of our treatment.  \S \ref{stokesetc}
introduces the Stokes parameters, Mueller matrix, and Jones vector and
matrix.  \S \ref{individualmatrices} defines the Mueller matrices for
the different effects we describe, including the mechanical rotation of
the feed with respect to the sky, the imperfections in the feed, the
feed, and the amplifier chain.  \S \ref{singlematrix} discusses the
combined effects of the receiver components and calculates that matrix
product.  \S \ref{matrixeval} describes the technique for evaluating the
matrix elements from observations.  \S \ref{applying} describes how to
apply the derived Mueller matrix to correct for the instrumental
effects. Finally, \S \ref{results} provides illustrative results for the
two 21-cm line receivers at Arecibo Observatory. This paper is a more
general and succinct version of Arecibo Technical and Operations Memo
2000-04, which covers all of Arecibo's receivers. 

\section{ STOKES, MUELLER, AND JONES} \label{stokesetc}

	The basic reference for our discussion of the fundamentals is
the excellent book on astronomical polarization by Tinbergen (1996).  A
more mathematical and fundamental reference is Hamaker, Bregman, and
Sault (1996), which the theoretically-inclined reader will find of
interest. In the following we make several unproven statements and
assertions about Mueller and Jones matrices; the explanations and
justifications can be found in the abovementioned references. 

\subsection{ The Stokes and Jones vectors}

	The fundamental quantities are the four Stokes parameters
$(I,Q,U,V)$, which we write as the 4-element Stokes vector 

\begin{eqnarray} 
\label{IQUV}
{\bf S} = \left[ 
\begin{array}{c} 
    I \\ Q \\ U \\ V \\
\end{array} 
\; \right] \; .
\end{eqnarray} 

\noindent The Stokes parameters are time averages of electric field
products; we use the terms ``voltage'' and ``electric field''
interchangeably because the radio telescope's feed converts one to the
other. The Jones vector represents the  fields as orthogonal linear
polarizations $(E_X, E_Y)$,  and considers them as complex to account
for their relative phase:

\begin{eqnarray} 
\label{EXEY}
{\bf J} = \left[ 
\begin{array}{c} 
    E_X \\ E_Y
\end{array} 
\; \right] \; .
\end{eqnarray} 

	Instructive special cases include pure linear and pure circular
polarization. Orthogonal linear polarizations are, obviously, $({\bf
J_X, J_Y)} = ([1,0], [0,1])$ (In the text, we write these vectors as
rows instead of columns for typographical purposes). Orthogonal circular
polarizations are $({\bf J_L, J_R}) =  \left( { [1,i] \over \sqrt 2},
{[i,1] \over \sqrt 2} \right)$, where $i = \sqrt{-1}$. Orthogonal
polarizations satisfy, for example, $J_X \overline{ J_Y} = 0$, where the bar
over a symbol indicates the complex conjugate and all products are time
averages.

	It is straightforward to relate the Stokes parameters to the
components of the Jones vector:

\begin{mathletters}
\label{Sdefinition1}
\begin{equation}
I = E_X \overline{ E_X} + E_Y \overline{ E_Y}
\end{equation}
\begin{equation}
Q = E_X \overline{ E_X} - E_Y \overline{ E_Y}
\end{equation}
\begin{equation}
U = E_X \overline{ E_Y} + \overline{ E_X} E_Y 
\end{equation}
\begin{equation}
iV = E_X \overline{ E_Y} - \overline{ E_X} E_Y \; .
\end{equation}
\end{mathletters}

\subsection{ The Mueller and Jones matrices}

	When the fields pass through some device, such as a feed or an
amplifier, they suffer amplitude and phase changes. These modify the
Stokes parameters. The Mueller matrix is the transfer function between
the input and output of the device:

\begin{equation}
\label{transfereqn}
{\bf S_{out}} = {\bf M \cdot S_{in}} \; .
\end{equation}

\noindent The Mueller matrix is, in general, a $4 \times 4$ matrix in
which all elements may be nonzero (but they are not all independent). In
the usual way, we write

\begin{eqnarray} 
\label{Mueller}
{\bf M} = \left[ 
\begin{array}{cccc} 
   m_{II}   & m_{IQ}   & m_{IU}   &  m_{IV} \\
   m_{QI}   & m_{QQ}   & m_{QU}   &  m_{QV} \\
   m_{UI}   & m_{UQ}   & m_{UU}   &  m_{UV} \\
   m_{VI}   & m_{VQ}   & m_{VU}   &  m_{VV} \\
\end{array} 
\; \right] \; .
\end{eqnarray} 

\noindent The matrix elements are just the partial derivatives, for
example

\begin{equation}
m_{VQ} =  { \partial V_{out} \over \partial Q_{in}} \biggr|_{I_{in}, U_{in}}
\; .
\end{equation}

	Every Mueller matrix has its Jones matrix counterpart; the Jones
matrix is the transfer function for the voltages. We defer further
discussion of Jones matrices until our treatment of three specific cases
of interest for radio astronomical systems. 

\section{ THE RADIOASTRONOMICAL RECEIVER COMPONENTS AND THEIR MATRICES}

\label{individualmatrices}

	In this section, we consider the Mueller matrices of devices
that are encountered by the incoming radiation on its way from the sky
to the correlator output. We consider three devices. The first device
encountered by the incoming radiation is the telescope (${\bf
M_{sky}}$), which mechanically rotates the feed with respect to the sky.
The second device is the feed, which we split into two parts.  The first
part (${\bf M_F}$) has the ability to change the incoming linear to any
degree of elliptical polarization; by design, feeds are intended to
produce either pure linear or pure circular, but in practice the
polarization is mixed, i.e.~elliptical.  The second part (${\bf
M_{IFr}}$) describes imperfections in such a feed, specifically the
production of nonorthogonal polarizations. The third device is the
amplifier chain (${\bf M_A}$).

	We describe the Jones matrix of the incoming radiation in linear
polarization, as in equation~\ref{EXEY}. However, the feed matrix ${\bf
M_F}$ can radically change the polarization state; for example, as we
shall see a dual-circular feed changes the order of the Stokes
parameters in equation~\ref{IQUV}.  Thus the signal voltages, after
going through ${\bf M_F}$, are not intuitively described as $(E_X,E_Y)$,
because the $(X,Y)$ connote linear polarization. Therefore, for the
voltages after the output of ${\bf M_F}$, we will use the symbols $(E_A,
E_B)$, or simply $(A,B)$, to emphasize the fact that the state of
polarization can be arbitrary.

	We assume that the remaining device matrices ${\bf M_{IFr}}$ and
${\bf M_A}$ produce closely-matched replicas of the input Stokes
parameters because, by design (hopefully!), the imperfections in the
feed are small.  We will retain only first-order products of these
imperfections. 

\subsection{ Mueller matrix relating the radio source to the receiver
input} \label{msky}

	Astronomical continuum sources tend to have linear polarization
but very little circular polarization; we assume the latter to be zero.
Moreover, we express the source polarization as a fraction of Stokes
$I$. Thus, 

\begin{eqnarray} 
\label{IQUVsrc}
{\bf S_{src}} = \left[ 
\begin{array}{c} 
    1 \\ Q_{src} \\ U_{src} \\ 0 \\
\end{array} 
\; \right] \; .
\end{eqnarray} 

	A linearly polarized astronomical source has Stokes
$(Q_{src},U_{src})$ defined with respect to the north celestial pole
(NCP). The source polarization is conventionally specified in terms of
fractional polarization and position angle with respect to the NCP,
measured from north to east. We have

\begin{mathletters}
\begin{equation}
Q_{src} = P_{src} \cos 2 PA_{src}
\end{equation}
\begin{equation}
U_{src} = P_{src} \sin 2 PA_{src}
\end{equation}
\begin{equation}
P_{src} = (Q_{src}^2 + U_{src}^2)^{1/2}
\end{equation}
\begin{equation}
PA_{src} = 0.5 \tan^{-1} \left( U_{src} \over Q_{src} \right)
\end{equation}
\end{mathletters}

	As we track a source with an alt-az telescope, the parallactic
angle $PA_{az}$ of the feed rotates on the sky.  $PA_{az}$ is defined to
be zero at azimuth 0 and increase towards the east; for a source near
zenith, $PA_{az} \sim az$, where $az$ is the azimuth angle of the
source.  The Stokes parameters seen by the telescope are $(Q_{sky},
U_{sky})$, and are related to the source parameters by

\begin{eqnarray} 
\label{skymatrix}
{\bf M_{SKY}} = \left[ 
\begin{array}{cccc} 
 1 &     0     &     0    & 0 \\
 0 &  \cos 2PA_{az} & \sin 2PA_{az} & 0 \\
 0 & -\sin 2PA_{az} & \cos 2PA_{az} & 0 \\
 0 &     0     &    0     & 1 \\
\end{array} 
\; \right] \; .
\end{eqnarray} 

\noindent The central $2 \times 2$ submatrix is, of course, nothing but
a rotation matrix. 

	For an equatorially-mounted telescope, $PA_{az}$ doesn't change
unless the feed is mechanically rotated with respect to the telescope.
This was the case with the NRAO 140-foot telescope\footnote{The National
Radio Astronomy Observatory is a facility of the National Science
Foundation operated under cooperative agreement by Associated
Universities, Inc.}, which was one of the several systems we used during
the genesis of this work. This relative rotation between feed and
telescope can produce unintended changes in the feed response to
astronomical sources, which makes such telescopes inherently less
accurate for polarization measurement. The 140-foot telescope is the
last of the great equatorial radio telescopes and with its recent
closure a detailed discussion of these matters has become irrelevant;
the remainder of the present paper considers alt-az telescopes
exclusively. 

\subsection{Mueller matrix for a perfect feed providing arbitrary
elliptical  polarization} \label{matrixa}

	The feed modifies the incoming voltages with its Jones matrix.
Suppose that the feed mixes incoming linear polarizations with arbitrary
phase and amplitude, keeping the total power constant and retaining
orthogonality; this makes it a perfect feed that responds to elliptical
polarization. Following Stinebring (1982) and  Conway and Kronberg
(1969), we write the transfer equation for the feed as

\begin{eqnarray}
\label{jonesmatrixf}
\left[
\begin{array}{c}
   E_{A,out}  \\
   E_{B,out}  \\
\end{array}
\; \right] = \left[
\begin{array}{cc}
  \cos \alpha              &    e^{i\chi}  \sin \alpha    \\
   -e^{-i\chi}\sin \alpha  &     \cos \alpha             \\
\end{array}
\; \right] \left[
\begin{array}{c}
   E_{X,in}  \\
   E_{Y,in}  \\
\end{array}
\; \right] \; .
\end{eqnarray}

\noindent The feed can completely alter the polarization state, so the
output Jones voltages are more intuitively described by subscripts
$(A,B)$ instead of $(X,Y)$, which connote linear polarization.  Here
$\alpha$ is the amount of coupling into the orthogonal polarization and
$\chi$ is the phase angle of that coupling. For example, for a native
linear feed $\alpha = 0$ and $\chi = 0$; for a native circular feed
$\alpha = 45^\circ$ and $\chi = 90^\circ$. Using this with
equation~\ref{Sdefinition1}, we find

\begin{eqnarray} 
\label{feedmatrixf}
{\bf M_F} = \left[ 
\begin{array}{cccc} 
 1 & 0 & 0 & 0 \\
 0 & \cos 2\alpha           & \sin 2\alpha\cos\chi               & \sin 2\alpha\sin\chi \\
 0 & -\sin 2\alpha\cos\chi & \cos^2\alpha-\sin^2\alpha\cos2\chi & -\sin^2\alpha\sin2\chi  \\
 0 & -\sin 2\alpha\sin\chi & -\sin^2\alpha\sin2\chi & \cos^2\alpha+\sin^2\alpha\cos2\chi  \\   
\end{array} 
\; \right] \; .
\end{eqnarray} 

\noindent Notice that in the right-bottom $3 \times 3$ submatrix,  the
off-diagonal transposed elements are of opposite sign for two of the
three pairs and the same sign for one. This is not an algebraic error! 

	Some instructive special cases include: 

	{\bf (1)} A dual linear feed: $\alpha=0$, $\chi=0$, and
$M_F$ is diagonal.

	{\bf (2)} A dual linear feed rotated $45^\circ$ with respect to
$(X,Y)$:  $\alpha=45^\circ$, $\chi=0$, and

\begin{eqnarray} 
\label{feedmatrixf2}
{\bf M_F} = \left[ 
\begin{array}{cccc} 
 1 & 0 & 0 & 0 \\
 0 & 0 & 1 & 0 \\
 0 & -1 & 0 & 0 \\
 0 & 0 & 0 & 1 \\
\end{array} 
\; \right] \; .
\end{eqnarray} 

\noindent As expected, this interchanges Stokes $Q$ and $U$, together
with a sign change as befits rotation.

	{\bf (3)} A dual linear feed rotated $90^\circ$ with respect to
$(X,Y)$:  $\alpha=90^\circ$, $\chi=0$, and

\begin{eqnarray} 
\label{feedmatrixf3}
{\bf M_F} = \left[ 
\begin{array}{cccc} 
 1 & 0 & 0 & 0 \\
 0 & -1 & 0 & 0 \\
 0 & 0 & -1 & 0 \\
 0 & 0 & 0 & 1 \\
\end{array} 
\; \right] \; .
\end{eqnarray} 

\noindent As expected, this reverses the signs of Stokes $Q$ and $U$.

	{\bf (4)} A dual circular feed:  $\alpha=45^\circ$, $\chi=90^\circ$,
and

\begin{eqnarray} 
\label{feedmatrixf4}
{\bf M_F} = \left[ 
\begin{array}{cccc} 
 1 & 0 & 0 & 0 \\
 0 & 0 & 0 & 1 \\
 0 & 0 & 1 & 0 \\
 0 & -1 & 0 & 0 \\
\end{array} 
\; \right] \; .
\end{eqnarray} 

\noindent The combination $(\alpha = 45^\circ,\chi = 90^\circ)$ 
permutes the order of the Stokes parameters in the output vector, making
it $(I,V,U,-Q)$.  $(E_A \overline{ E_A} - E_B \overline{ E_B})$ provides Stokes
$V$, instead of the $Q$ written in equation~\ref{Sdefinition1}; in other
words, it makes the feed native dual circular.  With respect to linear
polarization, $\alpha = 45^\circ$ has the same effect as in case {\bf
(2)}, namely to interchange $(Q,U)$ and change the sign of $Q$, because
it is equivalent to a feed rotation of $45^\circ$. 

	{\bf (5)} A dual-elliptical feed: arbitrary $\alpha$, $\chi =
90^\circ$. If $\chi = 90^\circ$ then  orthogonal linear inputs produce
orthogonal elliptical outputs with the ellipticity voltage ratio equal
to $\tan \alpha$ (see Tinbergen 1996, Figure 2.1). Thus $\alpha =
0^\circ$ passes the linear polarizations without modification, $\alpha =
90^\circ$ reverses the sign of the position angle, $\alpha = 45^\circ$
produces orthogonal circular polarizations, and other values of $\alpha$
produce orthogonal elliptical polarizations. 

\subsection {An important restriction: we set $\chi = 90^\circ$}
\label{ilrestrict} 

	Generally speaking, we prefer either pure linear or pure
circular feeds, i.e.~we prefer either the combination $(\alpha,\chi) =
(0^\circ, 0^\circ)$ or  $(\alpha,\chi) = (45^\circ, 90^\circ)$ in ${\bf
M_F}$. If a feed is designed to produce circular polarization with
$(\alpha,\chi) = (45^\circ, 90^\circ)$ and instead produces elliptical
polarization whose major axis is aligned with the $X$ direction, then
$\chi=90^\circ$ but $\alpha \ne 45^\circ$. If the ellipse is not aligned
with $X$, then $\chi \ne 90^\circ$, but this is equivalent to having
$\chi=90^\circ$ and physically rotating the feed. Thus, without loss of
generality, we can take $\chi=90^\circ$ (also see \S \ref{ilf}). This
leads to great simplification in ${\bf M_F}$, whose restricted form
becomes

\begin{eqnarray} 
\label{feedmatrixfr}
{\bf M_{Fr}} = \left[ 
\begin{array}{cccc} 
 1 & 0               & 0 &          0      \\
 0 & \cos 2\alpha    & 0 & \sin 2\alpha    \\
 0 & 0               & 1 & 0               \\
 0 & -\sin 2\alpha   & 0 & \cos 2\alpha    \\   
\end{array} 
\; \right] \; .
\end{eqnarray}

\subsection {An imperfect feed} \label{ilf}

	Again we follow Stinebring (1982) and Conway and Kronberg
(1969), and represent the imperfections of a feed by the Jones matrix

\begin{eqnarray}
\label{jonesmatrixilf}
\left[
\begin{array}{c}
   E_{X,out}  \\
   E_{Y,out}  \\
\end{array}
\; \right] = \left[
\begin{array}{cc}
   1                  &            \epsilon_1 e^{i\phi_1}    \\
\epsilon_2 e^{-i\phi_2}  &                 1                \\
\end{array}
\; \right] \left[
\begin{array}{c}
   E_{X,in}  \\
   E_{Y,in}  \\
\end{array}
\; \right] \; .
\end{eqnarray}

\noindent Here the $\epsilon$'s represent undesirable cross coupling
between the two polarizations; for example, this might be caused by the
two linear probes not being exactly $90^\circ$ apart.  The $\phi$'s are
the phase angles of these coupled voltages.  This equation assumes that
the feed is ``good'', meaning that we need retain only first-order terms
in $\epsilon$ (which makes the diagonal elements unity); however, for
the moment we allow the phases to be arbitrary. 

	After a little algebra, we find the matrix for the {\bf
I}mperfect {\bf F}eed to be

\begin{eqnarray} 
\label{feedmatrixilf}
{\bf M_{IF}} = \left[ 
\begin{array}{cccc} 
  1 & 0 & \Sigma \epsilon \cos & \Sigma \epsilon \sin \\
  0 & 1 & \Delta \epsilon \cos & \Delta \epsilon \sin \\
  \Sigma \epsilon \cos & -\Delta \epsilon \cos & 1 & 0 \\
  \Sigma \epsilon \sin & -\Delta \epsilon \sin & 0 & 1 \\
\end{array} 
\; \right] \; ,
\end{eqnarray} 

\noindent where $\Sigma \epsilon \cos = \epsilon_1 \cos \phi_1 +
\epsilon_2 \cos \phi_2$; $\Sigma \epsilon \sin = \epsilon_1 \sin \phi_1
+ \epsilon_2 \sin \phi_2$; $\Delta \epsilon \cos = \epsilon_1 \cos
\phi_1 - \epsilon_2 \cos \phi_2$; $\Delta \epsilon \sin = \epsilon_1
\sin \phi_1 - \epsilon_2 \sin \phi_2$. The imperfections in a good feed
are completely specified by four independent parameters.

	The central 4-element submatrix is a rotation matrix that
represents an error in position angle of linear polarization (its
diagonal elements are unity because of our first-order expansion in
$\epsilon$); in-phase mutual voltage coupling between the two probes
causes an apparent rotation.  Its off-diagonal elements $\Delta \epsilon
\cos$ are impossible to measure without calibration sources whose
position angles are accurately known.  Moreover, a small rotation can
also occur because of mechanical inaccuracy in mounting the feed.  In
practice, these problems make it impossible to separate this factor, so
one might as well assume it is equal to zero and make appropriate
adjustments to the position angle {\it ex post facto}. 

	Assuming $\Delta \epsilon \cos =0$ is consistent with a more
stringent assumption, namely that $\epsilon \equiv \epsilon_1 =
\epsilon_2$ and $\phi \equiv \phi_1 = \phi_2$ so that, also, we have
$\Delta \epsilon \sin =0$.  Physically, this is equivalent to assuming
that the coupling in a linear feed arises from the two probes being not
quite orthogonal and that the coupling in each has the same relative
phase.  In other words, it makes the correlated output non-orthogonal. 

	This assumption might seem to be too restrictive because, by
requiring $\phi_1 = \phi_2$, it eliminates the possibility of
imperfections inducing a change in the ellipticity of the polarization. 
However, it leads to no loss in generality of our treatment, because the
out-of-phase coupling case is included in the feed matrix ${\bf M_F}$. 
Our restricted case of an imperfect feed is described by just two
parameters:

\begin{eqnarray} 
\label{feedmatrixilfr}
{\bf M_{IFr}} = \left[ 
\begin{array}{cccc} 
  1 & 0 & 2 \epsilon \cos\phi & 2 \epsilon \sin\phi \\
  0 & 1 & 0 & 0 \\
  2 \epsilon \cos\phi & 0 & 1 & 0 \\
  2 \epsilon \sin\phi & 0 & 0 & 1 \\
\end{array} 
\; \right] \; .
\end{eqnarray} 

\subsection{The amplifiers}

	The two polarization channels go through different amplifier
chains. Suppose these have {\it voltage} gain $(g_A, g_B)$, {\it power}
gain $(G_A, G_B) = (g_A^2, g_B^2)$, and phase delays $(\psi_A, \psi_B)$.
The Jones matrix is

\begin{eqnarray}
\label{jonesmatrix2}
\left[
\begin{array}{c}
   E_{A,out}  \\
   E_{B,out}  \\
\end{array}
\; \right] = \left[
\begin{array}{cc}
   g_A e^{i \psi_A}    &            0                     \\
           0             &      g_B e^{i \psi_B}          \\
\end{array}
\; \right] \left[
\begin{array}{c}
   E_{A,in}  \\
   E_{B,in}  \\
\end{array}
\; \right] \; 
\end{eqnarray}

\noindent In practice, the amplifier gains and phases are calibrated
with a correlated noise source (the ``cal'').  Thus, our amplifier gains
$(G_A, G_B)$ have nothing to do with the actual amplifier gains. 
Rather, they represent the gains as calibrated by specified cal
intensities, one for each channel.  If the {\it sum} of the specified
cal intensities is perfectly correct, then the absolute intensity
calibration of the instrument is correct for an unpolarized source
(Stokes $I$ is correctly measured in absolute units).  In our treatment,
$G_A +G_B = 2$ by necessity because we deal with fractional
polarizations.  

	If the {\it ratio} of the cal intensities is correct, then the
difference between the two polarization channels is zero for an
unpolarized source. This happy circumstance does generally not obtain. 
However, the relative cal intensities are known fairly well, which
allows us to assume $\Delta G \equiv G_A - G_B \ll 1$ and to carry
$g_Ag_B$ to first order only, meaning we take $g_Ag_B=1$.  With this
first-order approximation, we have

\begin{eqnarray} 
\label{feedmatrixb}
{\bf M_A} = \left[ 
\begin{array}{cccc} 
  1                &  {\Delta G \over 2}   &     0    &      0          \\
{\Delta G \over 2}  &          1           &     0    &      0          \\
  0                &       0              & \cos \psi & -\sin \psi  \\
  0                &       0              & \sin \psi & \cos \psi  \\
\end{array} 
\; \right] \; .
\end{eqnarray} 

\noindent The incorrect relative cal amplitudes produce coupling of
Stokes $I$ into $Q$ through the nonzero $m_{QI}$; the difference between
the relative cal and sky phases produces a transfer of power between the
two correlated outputs, as we now discuss.

	The difference between the amplifier phases is also referred to
the cal.  Thus a phase difference $\psi = \psi_A - \psi_B$ represents
the phase difference that exists between a  linearly polarized
astronomical source and the cal and has nothing to do with the amplifier
chains. The behavior of this phase difference depends on the native
polarization of the feed.

	For a perfect native dual {\it linear} feed, the phase of a
linearly polarized astronomical source, modulo $180^\circ$, is
independent of $PA_{az}$ because the linearly polarized dipoles have no
relative phase difference ($\chi = 0$); the phase changes by $180^\circ$
when the measured $U$ changes sign as $PA_{az}$ changes.  Thus, $\psi$
modulo $180^\circ$ is independent of $PA_{az}$.  The desirable case
$\psi=0$ means that $(E_A \overline{ E_B} + \overline{ E_B} E_A)$ contains pure
Stokes $U$ and $(E_A \overline{ E_B} - \overline{ E_B} E_A)$ pure Stokes $V$. 

	For a perfect native dual {\it circular} feed, the phase of a
linearly polarized astronomical source (and therefore $\psi$) rotates as
$\pm2PA_{az}$ (see section~\ref{nearcirc} and
equation~\ref{paeqncirc2}).  At $PA_{az}=0$, the condition $\psi=0$
produces the correctly defined Stokes parameters in
equation~\ref{feedmatrixf4}, $(E_A \overline{ E_B} + \overline{ E_B} E_A) =
U_{sky}$ and $(E_A \overline{ E_B} - \overline{ E_B} E_A) = -Q_{sky}$.  If $\psi
\neq 0$, then the correctly defined Stokes parameters occur at
$PA_{az}=\pm {\psi \over 2}$. 

\subsection{ The correlator outputs}

	Our measured quantities are four time averaged voltage products
from the digital correlator: from autocorrelation, $(E_A E_A, E_B E_B)$;
from crosscorrelation, $(E_A E_B, E_B E_A)$. In these products we
consider the second quantity to be delayed relative to the first. Each
correlation function has $N$ channels of delay. We Fourier transform
(FT) these quantities to obtain spectra. 

	The autocorrelation functions are symmetric and thus their FT's
are real, with no imaginary components. We combine the two measured
crosscorrelation functions into a single one with $2N$ channels; it has
both negative and positive delays and is generally not symmetric, so its
FT is complex. We combine these Fourier transforms as in
equation~\ref{Sdefinition1}:

\begin{eqnarray} 
\label{Sdefinition}
\left[ 
\begin{array}{c} 
    A\! P\! B \\ A\! M\! B \\ A\! B \\ B\! A \\
\end{array} 
\; \right] = 
\left[ 
\begin{array}{c} 
    FT(E_AE_A) + FT(E_BE_B) \\  FT(E_AE_A) - FT(E_BE_B) \\ 
2 \; {\rm Re} [FT(E_AE_B)] \\ 2 \; {\rm Im} [FT(E_AE_B)]  \\
\end{array} 
\; \right] 
\; .
\end{eqnarray} 

\noindent If we have an ideal native dual linear feed and a perfect
receiver, then $(A\! P\! B, A\! M\! B, A\! B, B\! A)=(I, Q,U,V)_{sky}$; for native
circular, $(A\! P\! B, A\! M\! B,A\! B, B\! A) = (I,V,U,-Q)_{sky}$.

\section{ THE SINGLE MATRIX FOR THE RADIOASTRONOMICAL RECEIVER}

\label{singlematrix}

\subsection{ The general case with $\chi=90^\circ$} 

	The observing system consists of several distinct elements, each
with its own Mueller matrix.  The matrix for the whole system is the
product of all of them.  Matrices are not commutative, so we must be
careful with the order of multiplication.  

	We express the Jones vector of the incoming radiation in linear
polarization. The radiation first encounters the feed, producing Stokes
parameters as specified by ${\bf M_{Fr}}$ in
equation~\ref{feedmatrixfr}. Next it suffers the restricted set of
imperfections associated with ${\bf M_{IFr}}$
(equation~\ref{feedmatrixilfr}).  Finally it proceeds through the
amplifier chains, undergoing ${\bf M_A}$ (equation~\ref{feedmatrixb}). 
The product of these matrices, in this order (${\bf M_{TOT}} = {\bf M_A
\cdot M_{IFr} \cdot M_{Fr}}$), produces the vector that the correlator
sees, which we denote by ${\bf COR}$. In calculating ${\bf COR}$, we
ignore second order terms in the imperfection amplitudes $(\epsilon,
\Delta G)$ but, of course, retain all orders in their their phases
$(\phi, \psi)$ and also in the feed parameter $\alpha$. This gives

\begin{eqnarray} 
\label{totalmatrix}
{\bf M_{TOT}} = \left[ 
\begin{array}{cccc} 
                    1        & (-2 \epsilon \sin\phi \sin2\alpha + {\Delta G \over 2} \cos 2\alpha) &  2 \epsilon \cos\phi &  (2 \epsilon \sin\phi \cos2\alpha + {\Delta G \over 2} \sin 2\alpha) \\ 
{\Delta G \over 2}           &   \cos 2\alpha  &  0    &  \sin 2\alpha  \\
2\epsilon \cos(\phi + \psi)  & \sin 2\alpha \sin\psi & \cos\psi  &  -\cos2\alpha \sin\psi  \\
2\epsilon \sin(\phi + \psi)  & -\sin 2\alpha \cos\psi &  \sin \psi  &  \cos2\alpha \cos\psi 
\end{array}
\; \right] \; .
\end{eqnarray} 

\noindent The terms in the top row make $I \ne 1$ for a polarized
source. If one derives fractional polarization, for example $Q \over I$,
then it will be in error by amounts comparable to $[(\epsilon, \Delta G)
\times (Q,U,V)]$. For the weakly polarized sources we use as
calibrators, these products are second order and therefore are of no
concern. 

	However, for a strongly polarized source such as a pulsar,
these terms are first order. This can be particularly serious for
timing, because polarization variations across the pulse will produce
errors in the pulse shape. These effects can be eliminated by correcting
for the Mueller matrix. 

\subsection{ Two important special cases}

	Commonly, feeds are intended to be either pure linear or circular.
For these two important cases we have the following, expanded to first
order:

	{\bf (1)} A dual-linear feed with a slight elliptical component,
meaning that $[\alpha=(0^\circ, 90^\circ) + \delta\alpha]$ with $\delta
\alpha \ll 1$. We ignore second order products involving $\delta\alpha$:

\begin{eqnarray} 
\label{special1}
{\bf M_{TOT,lin}} = \left[ 
\begin{array}{cccc} 
  1                         &  \pm {\Delta G \over 2}     & 2\epsilon \cos\phi & 2\epsilon \sin\phi       \\
{\Delta G \over 2}          &  \pm 1                     &     0              & \pm 2\delta\alpha      \\
2 \epsilon \cos(\phi+\psi)  &  \pm 2\delta\alpha \sin\psi & \cos\psi           & \mp \sin\psi \\
2 \epsilon \sin(\phi+\psi)  &  \mp 2\delta\alpha \cos\psi & \sin\psi           & \pm \cos\psi  \\
\end{array} 
\; \right] \; .
\end{eqnarray} 

\noindent For terms with two signs, the top sign is for the $0^\circ$
case and the bottom for the $90^\circ$ case.

	{\bf (2)} As above, but for a dual-circular feed with $[\alpha =
(45^\circ, 135^\circ) + \delta\alpha]$:

\begin{eqnarray} 
\label{special2}
{\bf M_{TOT,circ}} = \left[ 
\begin{array}{cccc} 
  1                         &  \mp 2\epsilon \sin\phi   & 2\epsilon \cos\phi & \pm {\Delta G \over 2} \\
{\Delta G \over 2}          &  \mp 2\delta\alpha         &     0              & \pm 1      \\
2 \epsilon \cos(\phi+\psi)  &  \pm \sin\psi              & \cos\psi           & \pm 2\delta\alpha \sin\psi \\
2 \epsilon \sin(\phi+\psi)  &  \mp \cos\psi              & \sin\psi           & \mp 2\delta\alpha \cos\psi  \\
\end{array} 
\; \right] \; .
\end{eqnarray} 

\noindent Recall that in the circular case, the order of the Stokes
parameters in the output vector is permuted: $[I, V, U, -Q]$. As with the
imperfect linear, the imperfections in ${\bf M_{IFr}}$ produce coupling
between Stokes $I$ and $(V,U,-Q)$, represented by the nonzero elements in
the left column. The order of the terms in the column is independent of
the feed polarization, but the Stokes parameters are not, so the
imperfections produce different effects for the two types of feed.

\section {EVALUATING THE PARAMETERS IN THE MATRIX}

\label{matrixeval}

	We evaluate the parameters in equation~\ref{totalmatrix} using
observations of a polarized source tracked over a wide range of position
angle $PA_{az}$. The source is described by ${\bf S_{src}}$ and the
Mueller matrix for the radiation entering the feed by ${\bf M_{sky}}$,
both described in section~\ref{msky}.  The full Mueller matrix is ${\bf
M_{TOT} \cdot M_{sky}}$. The product of this matrix with ${\bf S_{src}}$
results in a set of of four equations, one for each element of the
observed ${\bf COR}$ vector. Recalling that we define the source Stokes
parameters as fractional (so that $I_{src}=1$) and that we assume
$V_{src}=0$, they are expressed by the four equations embodied in

\begin{eqnarray}
\label{paeqn}
\left[
\begin{array}{c}
   A\! P\! B  \\
   A\! M\! B  \\
   A\! B   \\
   B\! A  \\
\end{array}
\; \right] = {\bf M_{TOT} \cdot M_{SKY} \ \cdot}
\left[
\begin{array}{c}
   1               \\
   Q_{src}        \\
   U_{src}         \\
   0         \\
\end{array}
\; \right] \; .
\end{eqnarray}

	In practice, we cannot reliably measure the $PA_{az}$ dependence
of $A\! P\! B$.  $A\! P\! B$ is approximately equal to the total intensity, and it
is rendered inaccurate by small gain errors.  Thus, in practice we use
fractional correlator outputs.  We define

\begin{equation}
{\bf COR'} = { {\bf COR} \over A\! P\! B}
\end{equation}

\noindent and consider only the last three equations in
equation~\ref{paeqn}:

\begin{eqnarray}
\label{paeqnmod}
\left[
\begin{array}{c}
   A\! M\! B'  \\
   A\! B'   \\
   B\! A'  \\
\end{array}
\; \right] = {\bf M_{TOT} \cdot M_{SKY} \ \cdot}
\left[
\begin{array}{c}
   Q_{src}        \\
   U_{src}         \\
   0         \\
\end{array}
\; \right] \; .
\end{eqnarray}

\noindent Note that the division by $A\! P\! B$ produces errors in the other
elements of ${\bf COR'}$, but these errors are second order because they
are products of $\Delta G$ and/or $\epsilon$ with quantities such as
$Q_{src}$ that are already first order. Our whole treatment neglects
second order products, so we can neglect these errors. 

	Multiplying out equation~\ref{paeqnmod}, we obtain three
equations of the form 

\begin{equation}
\label{ambeqn}
A\! M\! B = A_{A\! M\! B'} + B_{A\! M\! B'} \cos 2PA_{az} + C_{A\! M\! B'} \sin 2PA_{az} \; , 
\end{equation}

\noindent where the coefficients $(A,B,C)$ are complicated function of
the 5 parameters $(\alpha, \epsilon, \phi, \Delta G, \psi)$ and, also,
the two source Stokes parameters $(Q_{src}, U_{src})$ which are not known
{\it ab initio}, so we have 7 unknown parameters. We have 9 measured
quantities, three $(A,B,C)$ for each correlator output; these are
derived from least squares fits of $(A\! M\! B',A\! B',B\! A')$ to $PA_{az}$. We use
nonlinear least squares fitting to solve for the 7 unknown parameters.
In practice, we use numerical techniques to obtain the relevant
derivatives. An alternative fitting technique that is useful when
combining the results from $N_{src}$ different sources into a grand
average is to express the coefficients $A_{A\! M\! B'}$, etc., in terms of the
$(5 + 2N_{src})$ unknown coefficients, and to lump all observations of
all sources together in one grand nonlinear least squares fit. 

	Nonlinear least square fitting is often plagued by multiple
minima, and the present case is no exception when the polarization is
nearly pure linear or circular. To discuss these cases we temporarily
assume $(\epsilon, \Delta G)=0$, which makes the mathematics more
transparent. 

\subsection{The nearly linear case}

	Equation~\ref{paeqn} becomes

\begin{mathletters}
\begin{eqnarray}
\label{paeqnlin1}
\left[
\begin{array}{c}
   Q_{out}  \\
   U_{out}  \\
   V_{out}  \\
\end{array}
\; \right] \approx \left[
\begin{array}{c}
   A\! M\! B'  \\
   A\! B'  \\
   B\! A'  \\
\end{array}
\; \right] = \left[
\begin{array}{cc}
 \pm 1                            &  0 \\
 \pm 2 \delta\alpha \sin\psi   &  \cos \psi  \\
 \mp 2 \delta\alpha \cos\psi   &  \sin \psi 
\end{array}
\; \right] \left[
\begin{array}{cc}
    \cos 2PA_{az} &  \sin 2PA_{az}  \\
   -\sin 2PA_{az} &  \cos 2PA_{az}
\end{array}
\; \right] \left[
\begin{array}{c}
   Q_{src}  \\
   U_{src}  \\
 \end{array}
\; \right] \; 
\end{eqnarray}

\noindent which, when multiplied out, becomes

\begin{eqnarray} \label{paeqnlin3} 
\tiny{
\left[ \begin{array}{c}
   A\! M\! B'  \\
   A\! B'  \\
   B\! A'  \\
\end{array}
\; \right] = \left[
\begin{array}{cc}
 \pm \cos 2PA_{src}  &   \pm \sin 2PA_{src}   \\
 \cos\psi \sin2PA_{src} \pm 2\delta\alpha \sin\psi \cos 2PA_{src}   &  -\cos\psi \cos 2PA_{src} \pm 2\delta\alpha \sin\psi \sin 2PA_{src} \\
 \sin\psi \sin2PA_{src} \mp 2\delta\alpha \cos\psi \cos 2PA_{src}   &  -\sin\psi \cos 2PA_{src} \mp 2\delta\alpha \cos\psi \sin 2PA_{src} 
\end{array}
\; \right] \left[
\begin{array}{c}
   P_{src} \cos 2PA_{az}  \\
   P_{src} \sin 2PA_{az}  \\
 \end{array}
\; \right] \; .
}
\end{eqnarray}

\noindent For $\delta \alpha=0$ we have

\begin{eqnarray}
\label{paeqnlin2}
\left[
\begin{array}{c}
   A\! M\! B'  \\
   A\! B'  \\
   B\! A'  \\
\end{array}
\; \right] = \left[
\begin{array}{cc}
 \pm \cos 2PA_{src}  &   \pm \sin 2PA_{src}   \\
 \cos\psi \sin 2PA_{src} &  -\cos\psi \cos 2PA_{src} \\
 \sin\psi \sin 2PA_{src} &  -\sin\psi \cos 2PA_{src} 
\end{array}
\; \right] \left[
\begin{array}{c}
   P_{src} \cos 2PA_{az}  \\
   P_{src} \sin 2PA_{az}  \\
 \end{array}
\; \right] \; .
\end{eqnarray}
\end{mathletters}

\noindent This shows that, for $\delta\alpha \rightarrow 0$, one cannot
distinguish between the two cases $(\alpha,\psi)=(0^\circ, \psi_0)$ and
$(\alpha,\psi)=(90^\circ, \psi_0 + 180^\circ)$, because the two bottom
rows change sign for $\psi_0 + 180^\circ$.  For these two cases, the
signs of $(Q,U)$ change, which is equivalent to rotating the feed by
$90^\circ$. The physical interpretation is straightforward:
$\alpha=90^\circ$ converts $E_X$ to $E_Y$ in
equation~\ref{jonesmatrixf}, thus changing the sign of $Q$; changing
$\psi$ by $180^\circ$ changes the sign of $U$; the combination is
equivalent to rotating the feed by $90^\circ$. 

	In practice one must deal with this problem.  For a conventional
linear feed, loosely described as two E-field probes in a circular
waveguide, the combination $(\alpha, \psi)=(90^\circ, 180^\circ)$ is
physically unreasonable.  However, for some feeds, in particular a
turnstile junction operating in linear polarization, either possibility
can occur and one must make the appropriate choice based on frequencies
well away from where the feed is pure linear; for a turnstile we expect
${d\alpha \over df} \approx const.$

\subsection{The nearly circular case} \label{nearcirc}

	Equation~\ref{paeqn} is

\begin{mathletters}
\begin{eqnarray}
\label{paeqncirc1}
\left[
\begin{array}{c}
   V_{out}  \\
   U_{out}  \\
   -Q_{out}  \\
\end{array}
\; \right] \approx \left[
\begin{array}{c}
   A\! M\! B  \\
   A\! B  \\
   B\! A  \\
\end{array}
\; \right] = \left[
\begin{array}{cc}
 \mp 2\delta\alpha                            &  0 \\
 \pm \sin\psi   &  \cos \psi  \\
 \mp \cos\psi   &  \sin \psi 
\end{array}
\; \right] \left[
\begin{array}{cc}
    \cos 2PA_{az} &  \sin 2PA_{az}  \\
   -\sin 2PA_{az} &  \cos 2PA_{az}
\end{array}
\; \right] \left[
\begin{array}{c}
   Q_{src}  \\
   U_{src}  \\
 \end{array}
\; \right] \; 
\end{eqnarray}

\noindent and becomes

\begin{eqnarray}
\label{paeqncirc2}
\left[
\begin{array}{c}
   A\! M\! B  \\
   A\! B  \\
   B\! A  \\
\end{array}
\; \right] = \left[
\begin{array}{cc}
 \mp 2\delta\alpha \cos 2PA_{src}  &   \mp 2\delta\alpha \sin 2PA_{src}   \\
 \sin(2PA_{src} \pm \psi)    &   -\cos(2PA_{src} \pm \psi)  \\
 \mp \cos(2PA_{src} \pm \psi)    &  \mp \sin(2PA_{src} \pm \psi)  \\
\end{array}
\; \right] \left[
\begin{array}{c}
   P_{src} \cos 2PA_{az}  \\
   P_{src} \sin 2PA_{az}  \\
 \end{array}
\; \right] \; .
\end{eqnarray}
\end{mathletters}

\noindent This shows that the angles $2PA_{src}$ and $\psi$ are
inextricably connected. We can determine only their sum (for
$\alpha=45^\circ$) or their difference. It also shows that the two
solutions $\alpha=(45^\circ, -45^\circ)$ are degenerate as $\delta
\alpha \rightarrow 0$, because the bottom row changes sign for these two
cases (thus flipping the derived sign of $Q_{src}$ and thereby rotating
the derived $PA_{src}$ by $90^\circ$). The physical interpretation of
this degeneracy is straightforward: for a pure circular feed, the phase
of a linearly polarized source rotates with $2PA_{az}$ and its absolute
value depends both on the system phase $\psi$ and the source position
angle $PA_{src}$. 

	In practice, one can only deal with this problem by having
additional information, namely knowing either $\psi$ or $PA_{src}$.  For
the case of turnstile junctions, which are narrow band devices, one can
determine $\psi$ at frequencies well away from that where the feed is
pure circular and interpolate.  Then, during the fit, this value of
$\psi$ should be fixed.  For a wideband circular feed, there is no
substitute for an independent calibration of the linearly polarized
position angle, either with a test radiator or with a source of known
polarization. 

\subsection{ Commentary}

	{\bf (1)} $\epsilon$ is the quadrature sum of the
$PA_{az}$-independent portions of the two correlated outputs $(A\! B,B\! A)$.
This power is distributed between those outputs according to $(\phi +
\psi)$. In the near-linear case, $\psi$ can change by $180^\circ$ by
changing the choice for $\alpha$, and this also produces a $180^\circ$
change in $\phi$. 

	{\bf (2)} Consider a high-quality standard linearly polarized feed
that has a correlated cal connected by equal-length cables.  Such a feed
has $\alpha \approx 0^\circ$ and the equal-length cables mean that $\psi
\approx 0^\circ$.  However, the solution yields $\psi \approx 180^\circ$
if the sign of $A\! M\! B$ is incorrect, which can easily happen if one
interchanges cables carrying the two polarizations; this is equivalent
to reversing the handedness of $PA_{az}$ and $PA_{src}$.  

	{\bf (3)} If one has a system without a correlated cal, then
$\psi$ is meaningless.  $\psi$ has contributions at r.f.  (from the
difference in cable and electrical lengths to the first mixer) and i.f. 
(from length differences after the mixer).  Normally the latter is
likely to dominate because the cable runs from the feed to the control
room are long.  For example, at Arecibo we found ${d\psi \over df} \sim
0.1$ rad MHz$^{-1}$, roughly constant among different systems.  This
corresponds to an electrical length difference of $\sim 5$ m, most of
which probably occurs along the pair of $\sim 500$ m optical fibers that
carry the two channels from the feed to the control room. 

	{\bf (4)} We have adopted the following procure for phase
calibration.  If there is a correlated cal then we measure $\psi_{cal}$
and fit it to a constant plus a slope $d\psi \over df$; we subtract this
fit from the source phase and produce corrected versions of $(A\! B,B\! A)$. 
Thus, the only component left in the correlated products is the
difference between source and cal phase, which is the same as $\psi$ in
the equation~\ref{feedmatrixb}. 

	If there is not a correlated cal, then we measure $\psi_{src}$
and fit it to a constant plus a slope $d\psi \over df$; we subtract the
{\it slope} but not the constant from the source phase and produce
corrected versions of $(A\! B,B\! A)$.  While most of this slope is in the
system, this procedure also subtracts away any intrinsic slope caused by
Faraday rotation. 

\section{ APPLYING THE CORRECTION} \label{applying}

	One of the major reasons to determine the Mueller matrix
elements is to apply it to observations and obtain true Stokes
parameters. There are two steps to this process.

\subsection{ Applying  ${\bf M_{TOT}}$ and ${\bf M_{sky}}$}

	The completely general form of equation~\ref{paeqn} uses 
observed voltage products instead of  fractional ones and does not force
$V=0$. Thus, to derive the source Stokes parameters from the data, we
use

\begin{eqnarray}
\label{paeqn1}
\left[
\begin{array}{c}
   I_{src}               \\
   Q_{src}        \\
   U_{src}         \\
   V_{src}         \\
\end{array}
\; \right] = {\bf (M_{TOT} \cdot M_{SKY})^{-1} \ \cdot}
\left[
\begin{array}{c}
   A\! P\! B  \\
   A\! M\! B  \\
   A\! B   \\
   B\! A  \\
\end{array}
\; \right] \; .
\end{eqnarray}

\subsection{ Deriving true astronomical position angles}

	The position angle of the source polarization $PA_{src}$ is
defined relative to the local idiosyncrasies.  For a dual linear feed
these include the angle at which feed probes happen to be mounted and,
also, which feed probe happens to be defined as $A$.  For a dual
circular feed this includes the phase angle at which the correlated cal
happens to be injected and the angle at which the feed happens to
be oriented. 

	Astronomers wish to express position angles in the conventional
way, {\it viz.}~with $PA_{src}$ measured relative to the North Celestial
Pole.  There is also the possibility of its handedness, but this is
taken care of automatically in the fitting process if the $PA_{az}$ is
correctly defined.  To satisfy the astronomers' desire, we must apply a
rotation matrix ${\bf M_{astron}}$, which looks like ${\bf M_{SKY}}$ in
equation~\ref{skymatrix} with $PA_{az}$ replaced by $\theta_{astron}$. 
For a linearly polarized feed the angle $\theta_{astron}$ is the angle
of the feed probes with respect to azimuth arm.  There is a sign
ambiguity that is best determined by empirical comparison with known
astronomical position angles.  At low frequencies, one must include the
effects of terrestrial ionospheric Faraday rotation, which is time
variable.

\section{ SAMPLE RESULTS}   \label{results}

	Here we present sample results for the two Arecibo L-band
receivers (L-band is the frequency range surrounding the 21-cm line). 
For each observing session the digital correlator was split into four
25-MHz chunks centered at different frequencies.  The L-band Wide
receiver (LBW) is a very wide-bandwidth dual linearly polarized feed. 
The L-band Narrow receiver (LBN) is a turnstile junction whose
polarization state changes from dual linear to dual circular over a
frequency range $\sim 100$ MHz.  First, however, we reiterate the
definitions of the parameters. 

\subsection{ Reiteration of parameter definitions}

	{\boldmath $\Delta G$} is the error in relative intensity
calibration of the two polarization channels. It results from an error
in the relative cal values $(T_{calA}, T_{calB})$. Our expansion
currently takes terms in $\Delta G$ to first order only, so if the
relative cal intensities are significantly incorrect then the other
parameters will be affected. 

	The relative cal values should be modified to make $\Delta G =
0$, keeping their sum the same. To accomplish this, make $T_{calA,
modified} = T_{calA} \left(1-{\Delta G \over 2}\right)$ and $T_{calB,
modified} = T_{calB} \left(1+{\Delta G \over 2}\right)$. 

	{\boldmath $\psi$} is the phase difference between the cal and
the incoming radiation from the sky; see the discussion following
equation~\ref{feedmatrixb}. It redistributes power between $(U,V)$ for a
dual linear feed and between $(Q,U)$ for a dual circular feed
(equations~\ref{special1} and \ref{special2}). 

	{\boldmath $\alpha$} is a measure of the voltage ratio of the
polarization ellipse produced when the feed observes pure linear
polarization. Generally, the electric vector traces an ellipse with
time; $\tan \alpha$ is the ratio of major and minor axes of the voltage
ellipse. Thus, $\tan^2 \alpha$ is the ratio of the powers. If a source
having fractional linear polarization $P_{src} = \sqrt{Q_{src}^2 +
U_{src}^2}$ is observed with a native circular feed that has $\alpha =
{\pi \over 4} + \delta \alpha$, with $\delta \alpha \ll 1$, then the
measured Stokes $V$ will change with $2PA_{az}$ and have peak-to-peak
amplitude $4 \delta \alpha$. 

	{\boldmath $\chi$} is the relative phase of the two voltages
specified by $\alpha$. Our analysis assumes $\chi = 90^\circ$; this
incurs no loss of generality, as explained in sections~\ref{ilrestrict} 
and \ref{ilf}. 

	{\boldmath $\epsilon$} is a measure of imperfection of the feed
in producing nonorthogonal polarizations (false correlations) in the two
correlated outputs. Our expansion takes $\epsilon$ to first order only. 
The only astronomical effect of nonzero $\epsilon$ is to contaminate the
polarized Stokes parameters $(Q,U,V)$ by coupling Stokes $I$ into them
at level $\sim 2 \epsilon$; the exact coupling depends on the other
parameters. For weakly polarized sources, this produces false
polarization; for strongly polarized sources such as pulsars, it also
produces incorrect Stokes $I$.

	{\boldmath $\phi$} is the phase angle at which the voltage
coupling $\epsilon$ occurs. It works with $\epsilon$ to couple $I$ with
$(Q,U,V)$. 

	{\boldmath $\theta_{astron}$} is the angle by which the derived
position angles must be rotated to conform with the conventional
astronomical definition.

\subsection{ Results for LBW}

	LBW is a very wide-band feed with native linear polarization. 
It has some problems with resonances.  Apart from these, the parameters
are nearly ideal and frequency-independent from 1175 to 1680 MHz:
$\alpha \sim 0.25^\circ$ and $\epsilon \sim 0.0015$.  Care was taken by
the receiver engineer to equalize the length of the two cables for the
correlated cal; as a result, $\psi \sim 4.6^\circ$. 


	Figure~\ref{lbwsrcplotfig} exhibits observational results for
two sources obtained during two Mueller-matrix calibration observing
sessions.  The digital correlator was split into four 25-MHz chunks,
with duplicate coverage at 1415 MHz on the two days; this provides seven
different center frequencies.  The change of position angle from Faraday
rotation is obvious over the broad band covered by the data, and even
within a single spectrum.  Regarding fractional polarization, the
decrease toward low frequency for B0017+154 is not surprising; this is
caused by Faraday depolarization.  However, that of B1634+269 reaches a
minimum near 1400 MHz, and this behavior is somewhat unusual.  We
believe that our measurements are correct and that this observed
behavior is real.

\subsection {Results for LBN}

	LBN is a turnstile system without a correlated cal and  is
commonly used over a large frequency range. Turnstile junctions are
narrow-band devices for two reasons: one, the polarization response is
defined by physical path length differences in waveguide, and these
length differences correspond to incorrect phase differences as one
moves away from the design frequency; two, unwanted reflections within 
the junction are eliminated by a tuning structure, and this is
narrow-band. The solid lines in Figure~\ref{lbnfig} exhibit the
frequency dependence of $\Delta G$, $\alpha$, $\epsilon$, and $\phi$ for
the 25 MHz bands centered at four frequencies within the range commonly
used with this feed. These particular data were derived from the source
B0017+154; we obtained data for two additional sources, and the results
agree well. The dashed lines are our adopted analytic expressions for
the frequency dependence of the parameters.

	Near 1415 MHz the polarization is dual circular; the absence of
a correlated cal means that the parameter $\psi$ has no meaning and,
moreover, we cannot measure the position angle of linear polarization. 
The dependence of $\alpha$ on frequency is close to linear, which is
what's expected for a turnstile junction.  The variation of $\epsilon$
is remarkably complicated, probably because of resonances in  the tuning
structure, and we do not have sufficient frequency coverage to
characterize it.  The scatter in $\phi$ for the 1375 MHz spectrum simply
reflects the uncertainty in the angle, which is large because $\epsilon$
is small.  Of course, $\Delta G$ simply reflects inaccurate relative cal
values and not the properties of the turnstile itself. 


\acknowledgements

	This work was supported in part by NSF grant 95-30590 to CH.

\clearpage

\section{FIGURE CAPTIONS}

\figcaption{Polarizations of B1634+269 and B0017+154, derived from the
calibration measurements for LBW. Position angles have not been rotated
by $\theta_{astron}$.  \label{lbwsrcplotfig} }

\figcaption{Mueller matrix parameters versus frequency for LBN, together
with our adopted analytic approximations. $\psi$ is meaningless and not
shown because this receiver does not have a correlated cal.
\label{lbnfig}}

\clearpage

\begin{figure}
\plotone{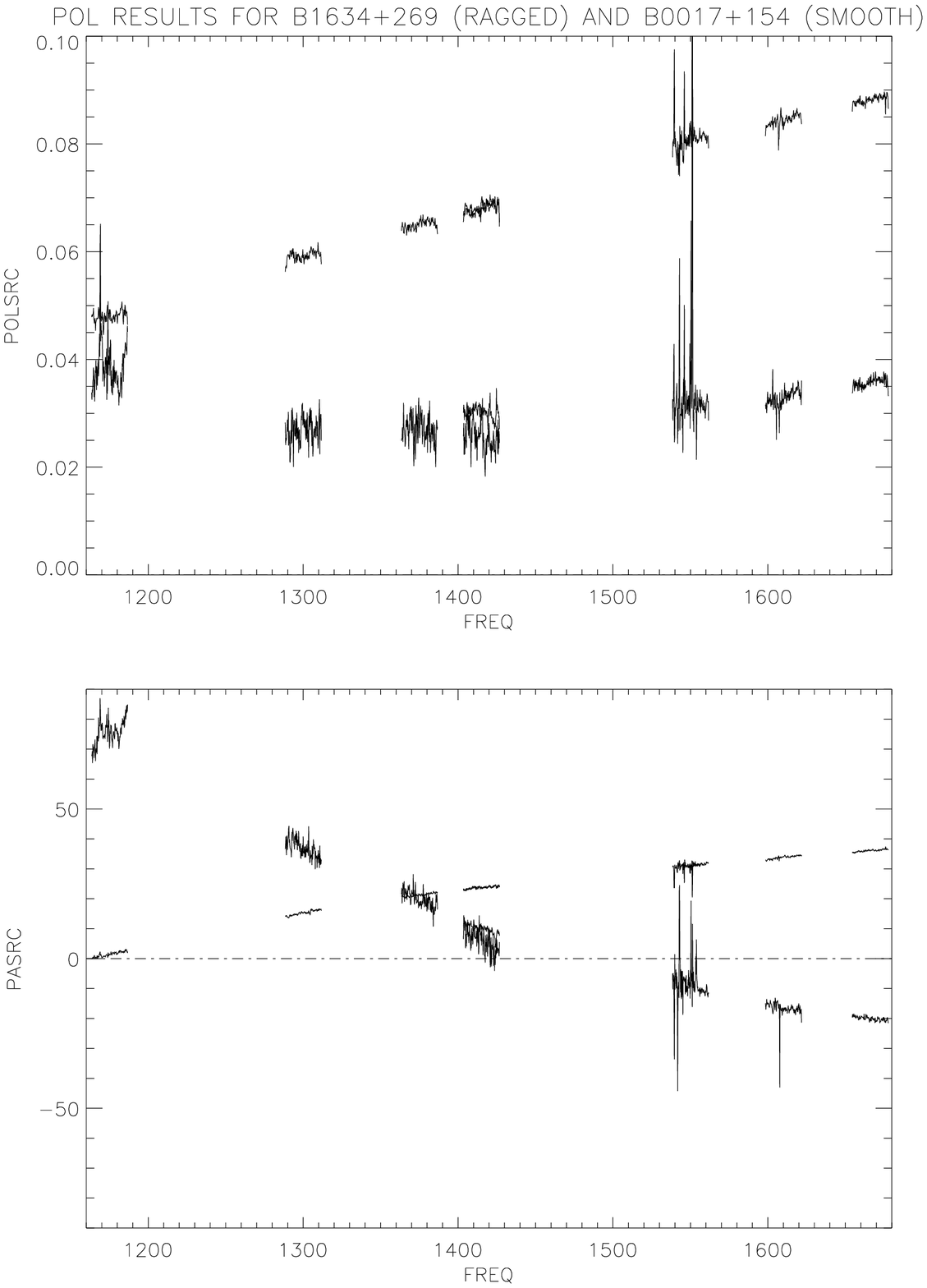}
\end{figure}

\clearpage

\begin{figure}
\plotone{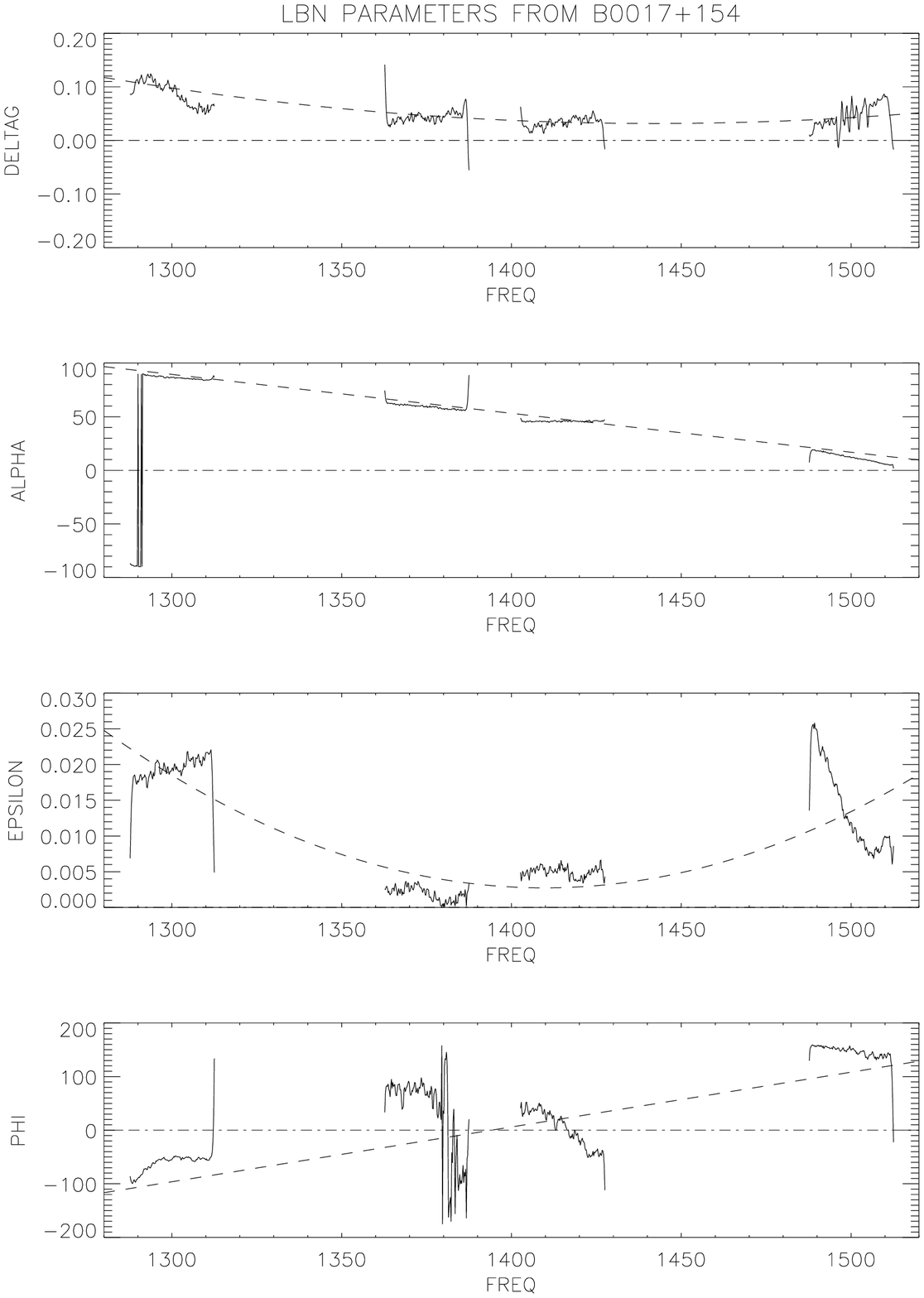}
\end{figure}

\enddocument
\end